\documentclass[namedreferences]{solarphysics}
\usepackage[hyperref,optionalrh,solaromanenum]{spr-sola-addons} 
\usepackage{graphicx}                    
\usepackage{rotating}
\usepackage{color}                       

\usepackage{breakurl}                         
\usepackage{booktabs}

\begin{document}

\begin{article}

\begin{opening}

\title{A Statistical Comparison of EUV Brightenings Observed by SO/EUI with Simulated Brightenings in Non-potential Simulations}

%

%


\author[addressref={aff1,aff2},corref,email={krzysztof.barczynski@pmodwrc.ch}]{\inits{K.}\fnm{Krzysztof}~\lnm{Barczynski}\orcid{0000-0001-7090-6180 }}
\author[addressref={aff6}]{\inits{K. A.}\fnm{Karen A.}~\lnm{Meyer}\orcid{0000-0001-6046-2811}}
\author[addressref={aff2,aff1}]{\inits{L.K.}\fnm{Louise K.}~\lnm{Harra}}
\author[addressref={aff5}]{\inits{D. H.}\fnm{Duncan H.}~\lnm{Mackay}}
\author[addressref={aff3}]{\inits{F.}\fnm{Frédéric}~\lnm{Auchère}}
\author[addressref={aff4}]{\inits{D.}\fnm{David}~\lnm{Berghmans}}

\address[id=aff1]{ETH-Zurich, Wolfgang-Pauli-Str. 27, 8093 Zurich, Switzerland}

\address[id=aff2]{Physikalisch-Meteorologisches Observatorium Davos, World Radiation Center, 7260, Davos Dorf, Switzerland}

\address[id=aff6]{Mathematics, School of Science \& Engineering, University of Dundee, Nethergate, Dundee, DD1 4HN, UK}

\address[id=aff5]{School of Mathematics and Statistics, University of St Andrews, North Haugh, St Andrews, KY16 9SS, UK}

\address[id=aff3]{Institut d'Astrophysique Spatiale, CNRS, Univ. Paris-Sud, Universite Paris-Saclay, Bat. 121, 91405 Orsay, France}

\address[id=aff4]{Solar-Terrestrial Centre of Excellence – SIDC, Royal Observatory of Belgium, Ringlaan -3- Av. Circulaire, 1180 Brussels, Belgium}

\begin{abstract}
The High Resolution Imager (HRI$_{\rm EUV}$) telescope of the Extreme Ultraviolet Imager (EUI) instrument  onboard Solar Orbiter has observed 
EUV brightenings, so-called campfires, as fine-scale structures at coronal temperatures.
The goal of this paper is to compare the basic geometrical (size, orientation) and physical (intensity, lifetime) properties of the EUV brightenings with regions of energy dissipation in a non-potential coronal magnetic field simulation.
In the simulation, HMI line-of-sight magnetograms are used as input to drive the evolution of solar coronal magnetic fields and energy dissipation.
We applied an automatic EUV brightening detection method to EUV images obtained on 30 May 2020 by the HRI$_{\rm EUV}$ telescope.
We applied the same detection method to the simulated energy dissipation maps from the non-potential simulation to detect simulated brightenings.
We detected EUV brightenings with density of  $1.41\times10^{-3}$ brightenings/Mm$^2$ in the EUI observations and simulated brightenings between $2.76\times 10^{-2}$ - $4.14\times 10^{-2}$ brightenings/Mm$^2$ in the simulation, for the same time range.
Although significantly more brightenings were produced in the simulations, the results show similar distributions of the key geometrical and physical properties of the observed and simulated brightenings.
We conclude that the non-potential simulation can successfully reproduce statistically the characteristic properties of the EUV brightenings (typically with more than 85\% similarity); only the duration of the events is significantly different between observations and simulation. Further investigations based on high-cadence and high-resolution magnetograms from Solar Orbiter are under consideration to improve the agreement between observation and simulation.
\end{abstract}

%

\end{opening}

\section{Introduction}\label{s:intro}
EUV brightenings, so-called campfires, are small-scale (0.4 - 4 Mm), short-lived (10 - 200 s) brightenings observed in the quiet Sun, at Extreme Ultraviolet (EUV) wavelengths and coronal temperatures \citep{Berghmans2021}.
The EUV brightenings were observed with the High-Resolution Imager (HRI) - one of the three telescopes in the Extreme Ultraviolet Imager (EUI) instrument onboard the Solar Orbiter mission.
The EUV brightenings are observed in the quiet Sun as either small-scale loop-like, dot-like, or complex features.
The comparison with simultaneous SDO/AIA observations shows that most of the EUV brightenings can be identified in 171~\AA, 193~\AA, 211~\AA,~and 304\AA~\citep{Berghmans2021}.
However, the nature of the EUV brightenings is an open question.

Previous high-resolution observations, obtained by the Hi-C rocket \citep{Kobayashi2014, Rachmeler2019}, detected small-scale loop-like structures \citep{Peter2013, Barczynski2017} in the plage region.
These miniature loops have geometrical properties and lifetimes similar to EUV brightenings observed with EUI.
Previous EUI observations have shown that the EUV brightenings are located between 1000 km and 5000 km above the photosphere \citep{Zhukov2021}.
 Most EUV brightenings appear to be located at the neutral line between patches of two opposite magnetic field polarities \citep{Panesar2021, Kahil2022}, indicating the importance of the magnetic field in the formation and evolution of these features.

 Short-lived, small-scale structures ($<$ 5Mm), such as EUV brightenings, are numerous in the transition region and the solar corona. 
Despite their small size and short lifetime, their large number can significantly influence the physical processes in the transition region and the solar corona. 

\citet{Meyer2013}  found numerous small-scale brightenings in simulations of the solar atmosphere before the first HRI$_{\rm EUV}$ telescope measurements of EUV brightenings were made. The simulations showed small, short-lived regions of energy dissipation in non-linear force-free field simulations of the Sun's small-scale corona, with input constraints from quiet Sun magnetograms from the Helioseismic and
Magnetic Imager \citep[HMI:][]{Scherrer2012} onboard the Solar Dynamics Observatory \citep[SDO:][]{Pesnell2012}. The energy dissipation was found to be largest low down in the simulation, close to the photosphere.
Using the same simulation method as \cite{Meyer2013}, we investigate the properties of the simulated brightenings and compare them statistically with the new EUI observations from Solar Orbiter.

%

In this paper, we compare the EUV brightenings observed by HRI (Section~\ref{s:obs}) with simulated brightenings obtained from the non-potential simulation (Section~\ref{s:sym}) using statistical methods.
We discuss both the similarities and differences of the observed and simulated brightenings (Section~\ref{s:cf_comp}).
Finally, we summarise our results, and anticipate even higher resolution results available in the future (Section~\ref{s:conclusion}).

\section{Observations}\label{s:obs}

\subsection{EUI Observations}\
We used the solar atmosphere images obtained with the Extreme Ultraviolet Imager \citep[EUI:][]{Rochus2020} onboard Solar Orbiter \citep{Muller2020}.
The EUI instrument consists of three telescopes: the dual-band Full-Sun Imager (FSI) working at~174~\AA~and~304~\AA; the High-Resolution Imager  observing in the hydrogen Lyman-$\alpha$ line (HRI$_{\rm Ly-\alpha}$) and the High-Resolution Imager observing in EUV at 174~\AA~(HRI$_{\rm EUV}$).
We analysed level-2 data\footnote{EUI Data Release 4, \url{doi.org/10.24414/s5da-7e78}} from HRI$_{\rm EUV}$.
The emission observed with HRI$_{\rm EUV}$ is dominated by Fe IX and Fe X lines and corresponds to the upper transition region and lower corona temperature (1~MK). 

We studied a sequence of 50 images of the quiet Sun region obtained on 30 May 2020 between 14:54:00 UTC and 14:58:05 UTC with 5~s cadence and exposure time of 3~s.
During the observation, Solar Orbiter was 0.556 AU from the Sun and had an angular separation of 31.5\textdegree~in solar longitude from the Earth-Sun line.
The images were projected to Carrington coordinates with a pixel size of 0.01625 heliographic degrees, which corresponds to 197 km.
The projection method is described by \cite{Berghmans2021}, where the field-of-view is 2400$\times$2400 pixels.

\subsection{HMI Data}\label{s:hmi}
 We used line-of-sight (LOS) magnetograms obtained with SDO/HMI \citep{Scherrer2012} as input data to simulate the solar coronal magnetic field.
HMI provides full-disk LOS magnetograms with a cadence of 45~s and pixel size corresponding to 364 km in the solar photosphere.
The exposure time of HMI magnetograms was 150 ms on 30 May 2020. \footnote{The HMI magnetogram exposure increases with time due to instrument degradation. The typical exposure time was in range 115-140 ms in years 2010-2015 \citep{Hoeksema_2018}. The information about exposure time is available at: \url{jsocstatus.stanford.edu/hk/long_term_trending/hmi/mechanisms.html}}

We analysed LOS magnetograms obtained on 30 May 2020 between 08:57:14 and 15:12:14 UTC, with a total of 501 magnetograms.
%
%
%
We used SDO/HMI pre-processed hmi.M\_45s data provided by the Joint Science Operations Center (JSOC: \url{jsoc.stanford.edu}). The data were derotated during JSOC pre-processing. 

A region of size $512\times512$ pixels ($\approx186\times186$ Mm) was selected within the HMI data. This region was selected as it overlaps with the EUI field of view and is close to flux balance, which is a requirement for the non-potential coronal simulation (Section~\ref{s:cmodel}). The magnetograms were smoothed temporally by averaging over nine frames, to remove five minute oscillations. The noise in the dataset was estimated to be $\sigma_B=6.2$ G, where $\sigma_B$ is the half-width half-maximum of a Gaussian fit to a histogram of pixel values. Pixels of magnitude less than $2\,\sigma_B$ were set to zero. To obtain exact flux balance the magnetograms were then corrected as follows: for each frame, the average imbalance $B_{i}$ per pixel was determined for all pixels of magnitude $\ge3\,\sigma_B$, then $B_{i}$ was subtracted from all such pixels. The average and maximum imbalance per pixel across all frames was $1.3$ G and $2.6$ G, respectively, so no pixels changed sign as a result of the correction.
It should be noted that while HMI data are used in the present study to drive the non-potential coronal simulation, the data mostly resolve magnetic fields on the scale of supergranules and do not match the higher resolution of the EUI observations.
We are however restricted to use HMI data as they were the highest spatial and temporal magnetogram data available to us at the time of the EUI observations. In future this restriction will be removed when SO/PHI data become available.

\subsection{EUV Brightenings Detection Method}

We used the automated EUV brightenings detection method presented by \citet{Berghmans2021}, with the same setup. The detection is made using a dyadic `\`a trous' wavelet transform using a $B_3$ spline scaling function \citep[e.g., ][]{Starck1994, Starck2002}. Using the treatment of \cite{Murtagh1995}, coefficients in the first two scales are considered significant when they are 5 times the standard deviation of the photon shot noise. The detection in individual images results in an ($x$, $y$, $t$) binary cube of events, for each of which geometrical and photometrical properties are computed. The surface area of an event is given by the total number of pixels of its projection along the temporal axis. Estimates of the length, width and orientation of each event are obtained by respectively the major axis, minor axis, and angle of a fitted ellipse. The total intensity is calculated as the integrated intensity during the EUV brightening duration. The relative variance of intensity is defined as the variance of the mean intensity at each time-step, normalized to the mean intensity of the event. Finally, the EUV brightening volume is defined as the number of ($x$, $y$, $t$) voxels.

\subsection{Detected EUV Brightenings}
This work aims to compare the EUV brightenings from observations with regions of energy dissipation in non-potential simulations to determine if the geometric and physical characteristics show any similarity or not.
The EUV brightenings detection method is sensitive to the spatio-temporal data resolution.
The EUI/HRI$_{\rm EUV}$ images have higher spatial and temporal resolution than HMI magnetograms that are used as the simulation input.
To adjust for this difference in spatial scales, we reduced the spatial resolution of EUI/HRI$_{\rm EUV}$ images to the spatial resolution of HMI and analysed only these six EUI/HRI$_{\rm EUV}$ images, which temporally correspond to six HMI magnetograms used in the simulation.

\begin{figure} 
 \centerline{\includegraphics[width=1.0\textwidth,clip=]{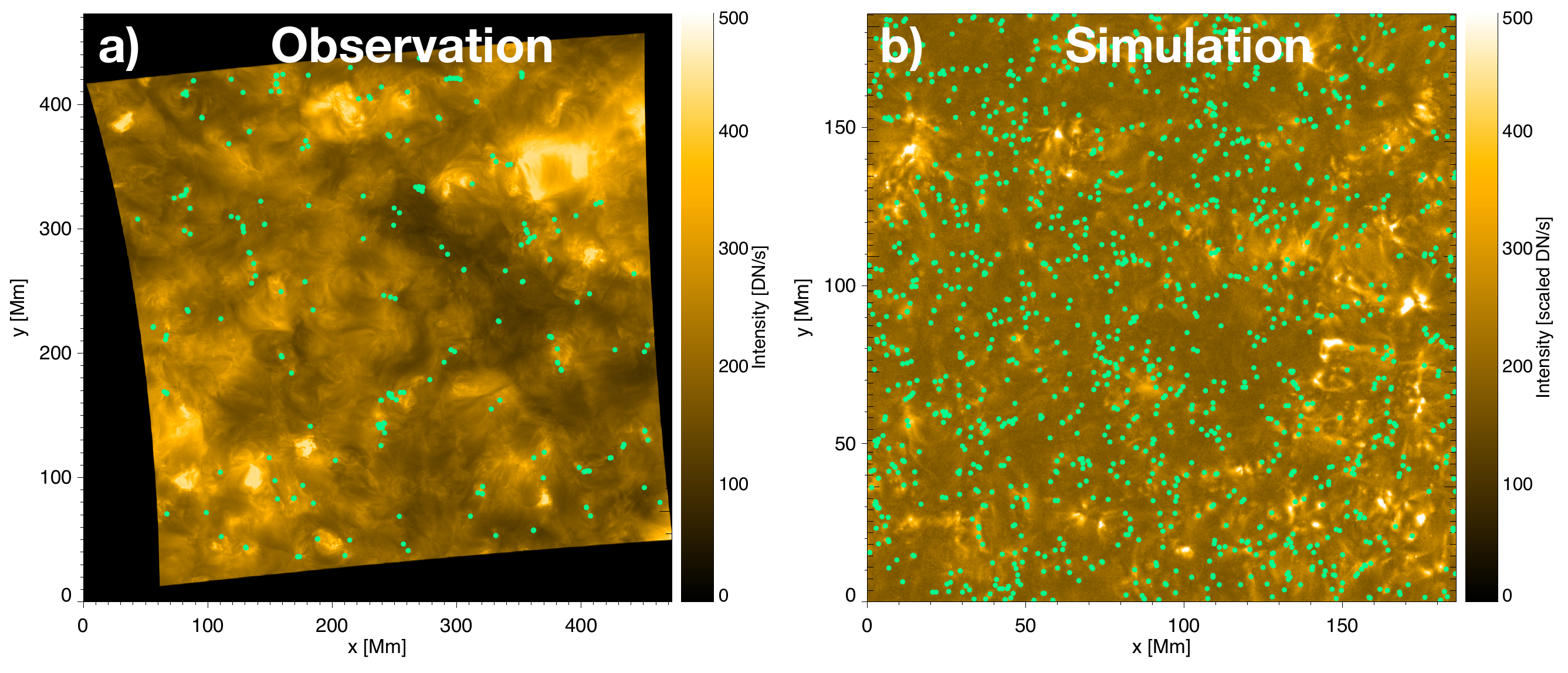}}
 \caption{Locations of brightenings (green dots) shown for (a) the observations and (b) the simulation intensity maps. Panel (a) shows the average of the Carrington projected HRI$_{\rm EUV}$ full-field of view image at 174~\AA~wavelength. Panel (b) shows the average map of the scaled simulated intensity. Both average maps were created from six temporally corresponding images.}\label{fig:imgs}
 \end{figure}

Then, we applied the automated EUV brightenings detection method \citep{Berghmans2021} which detected 240 EUV brightenings in EUI/HRI$_{\rm EUV}$ observations.
We define the brightenings density as the number of brightenings per area unit per observation time (4 min 30~s).
The EUV brightenings density is $1.41\times 10^{-3}$ brightenings/Mm$^2$. 
In Figure~\ref{fig:imgs}(a), we present the location of detected EUV brightenings in the average HRI$_{\rm EUV}$ intensity map.
The EUV brightenings are distributed non-uniformly in the intensity map.
They tend to form together in groups or concentrations to form elongated shapes.
Moreover, the saturated structures in  HRI$_{\rm EUV}$ images are not identified with detection algorithm as EUV brightenings.

%

\section{Simulation}\label{s:sym} 

\subsection{Coronal Model}\label{s:cmodel}
The prepared HMI magnetogram series was used directly as a lower boundary condition to drive the evolution of the simulated coronal magnetic field. The simulation domain was chosen to be $512\times512\times256$ grid cells ($\approx186\times186\times93$ Mm), which is periodic in the $x$ and $y$ directions, and closed at the top boundary. The initial condition for the simulation was a potential magnetic field extrapolated from the first  magnetogram frame at 08:57:14 UTC. A magnetofrictional method \citep{Yang1986} was used to evolve the coronal magnetic field through a continuous series of quasi-static non-linear force-free equilibria  \citep[e.g.][]{vanBalle2000}. The method has previously been successfully applied to observed magnetograms to simulate the coronal magnetic field evolution of the quiet Sun \citep[e.g.][]{Meyer2013,Madjarska2022} and active regions \citep[e.g.][]{Mackay2011, Gibb2014, Yardley2021}.

The magnetic field $\mathbf{B}=\nabla\times\mathbf{A}$ was evolved via the induction equation,
\begin{equation}\label{eqn:ind}
\frac{\partial\mathbf{A}}{\partial t}=\mathbf{v}\times\mathbf{B},
\end{equation}
where $\mathbf{A}$ is the vector potential. The magnetofrictional velocity $[\mathbf{v}]$, is defined as
$$\mathbf{v}=\frac1\nu\frac{\mathbf{j}\times\mathbf{B}}{B^2},$$
where $\nu$ is the coefficient of friction. The coefficient of friction is determined by the grid resolution ($\Delta x=364$ km) and time step (taking 450 relaxation steps between HMI magnetograms, $\Delta t = 0.1$ s), such that $\nu^{-1}=0.1(\Delta x)^2/\Delta t\approx1.3\times10^5$ km$^2$ s$^{-1}$.

We considered the simulated coronal magnetic field between 14:57:48 UTC and 15:01:54 UTC as this time period coincides with the EUI observations. The simulation started 6 hours before this, to allow sufficient time for the coronal magnetic field to evolve away from its initial potential field state.
Following the method of \citet{Meyer2012} and \citet{Meyer2013}, we computed the energy that is continually dissipated within the simulation due to the relaxation processes as
\begin{equation}\label{eqn:q}
    Q(x,y,z)=\frac{B^2}{4\pi}(\nu|\mathbf{v}^2|).
\end{equation}
We compare the properties of the energy dissipation within the simulation with the EUV brightening observations by integrating $Q$ along the line of sight:
\begin{equation}\label{eqn:qxy}
E_q(x,y)=\int_{z_\textrm{\tiny min}}^{z_\textrm{\tiny max}}Q(x,y,z)\,dz,
\end{equation}
where $z_\textrm{\tiny min}=0$ km is the base (photosphere) and $z_\textrm{\tiny max}=93~000$~km is the top boundary of the simulation domain. While the energy dissipation is integrated over the full vertical extent of the computational domain, the majority of the energy dissipated occurs low down in the computational domain.

The intensity of $E_q$ was then scaled to match the intensity found in the EUV brightening observations (see Section~\ref{s:scl}). As well as the simulation described above, four additional simulations were run to test whether any of the following had a significant effect on the results: the simulation start time; the inclusion of a diffusive term in Equation~(\ref{eqn:ind}); and an open top boundary condition. These simulations are described in Appendix~\ref{a:sim}, where the results were not significantly different to those presented in Section~\ref{s:cf_comp}.


\subsection{Intensity Scaling}\label{s:scl}
We focused on six frames obtained from the non-potential simulation that covered the EUI observation time. 
We took into account the correction of 228.6~s for the light travel time difference
between the Sun and Solar Orbiter and from the Sun to SDO/HMI.

\begin{figure} 
 \centerline{\includegraphics[width=1.0\textwidth,clip=]{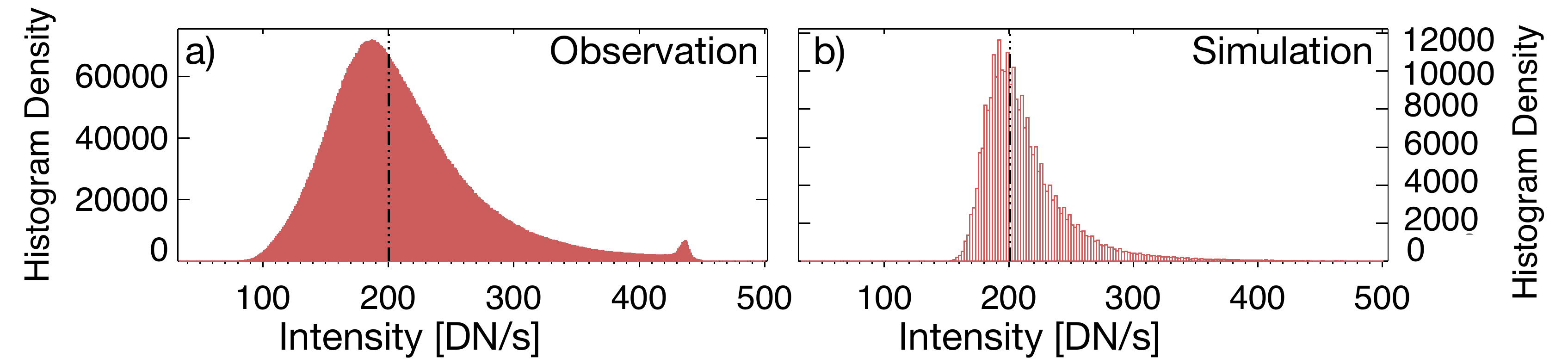}}
 \caption{Histogram intensity of (a) the observations and (b) the simulation. The histograms are based on 6 images. The dot-dashed line shows the median value.}\label{fig:histo_calib}
 \end{figure}

We scaled the six simulated frames which are co-temporal with the six EUI images.
We converted the intensity of the EUI observation from DNs to photons ($M_{\rm EUI}$) using the conversion factor 6.34375 DNs photon$^{-1}$.
We used a numerical method to scale the simulated ($M_{\rm SIM}$) data to have the same photon variance and the same median value as EUI data.
If $M_{\rm SIM, PH}$ is the scaled simulated data with added photon noise, then the scaling conditions are defined as:
     $$\mathrm{variance}(M_{\rm EUI}) = \mathrm{variance}(M_{\rm SIM, PH})$$
     $$\mathrm{median}(M_{\rm EUI}) = \mathrm{median}(M_{\rm SIM, PH})$$
To fulfil the above conditions, we introduced two scalar coefficients which we obtained numerically.
First, the variance scaling coefficient [$s$], scales the simulated data by the factor of $s$. %
The second coefficient [$A$] shifts the median value of simulated data.
The relationship between the original simulated data and scaled simulated data is given by:
$$M_{\rm SIM, PH} = s*M_{\rm SIM}+A+\mathrm{photon\_noise}(s*M_{\rm SIM}+A)$$
The photon noise is added to the scaled and shifted data. 
Subsequently, the EUI and scaled simulated data have the same photon variance and the same median value of photon number.
Finally, the EUI and scaled simulated data are converted from photon numbers to DNs.
Figure~\ref{fig:imgs}(b) shows the scaled simulated intensity alongside an EUI observation at the same time, while Figure~\ref{fig:histo_calib} shows the histograms of EUI intensity (panel a) and the scaled simulated intensity (panel b).

\subsection{Simulated Brightenings Detection}

We applied the automated EUV brightenings detection method \citep{Berghmans2021} to the scaled and shifted simulated intensity images under the same conditions as for the detection in  EUI/HRI$_{\rm EUV}$.
The algorithm found 1310 simulated brightenings, which corresponds to a density of $3.79\times 10^{-2}$ brightenings/Mm$^2$.
In Figure~\ref{fig:imgs}(b), we present the location of detected simulated brightenings in the average simulated intensity map.
The simulated brightenings also concentrate in groups.

\begin{figure} 
 \centerline{\includegraphics[width=1.0\textwidth,clip=]{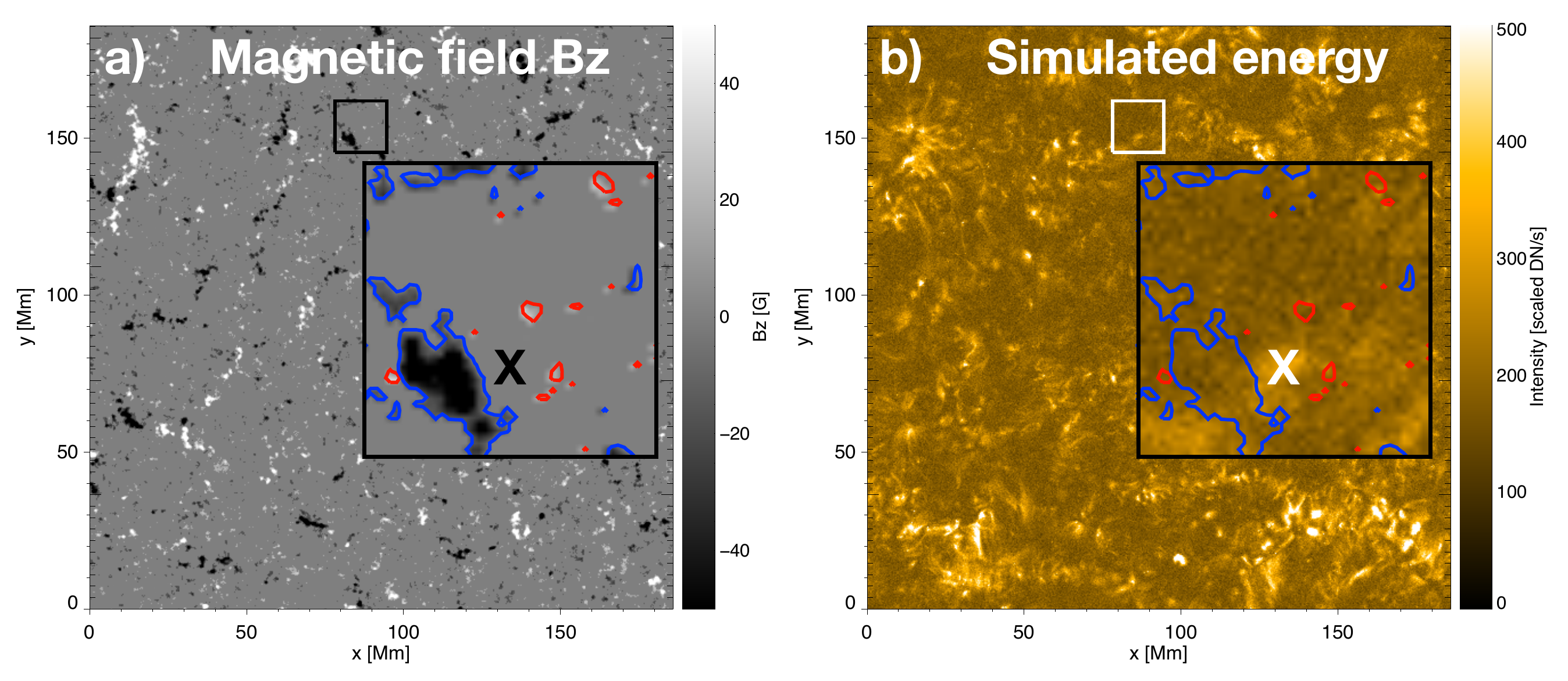}}
 \caption{The corresponding field of view of (a) the simulated magnetic field and (b) the simulated intensity. Panel (a) shows the Carrington projected B$_{z}$ magnetogram. Panel (b) shows the map of the scaled simulated intensity. Both average maps were created from temporally corresponding data (15:00:06 UT). The small-box marks the area zoomed in the large box. The contours show the magnetic field at level $\pm$10 G. The ``X" indicates the position of a simulated brightening located between patches of opposite polarity magnetic field.}\label{fig:mf_img}
 \end{figure}

Examples of simulated brightenings occurring between patches of opposite polarity magnetic field can be found by comparing the HMI magnetogram data at the lower boundary of the simulation with the simulated intensity maps. Figure~\ref{fig:mf_img}(a) shows the HMI magnetogram at 15:00:06 UT (with cleaning applied as described in Section~\ref{s:hmi}) and (b) shows the scaled, simulated intensity map at the same time. The small box on each image indicates the location of the zoomed region shown in the larger box. The ``X'' indicates the location of one of the detected simulated brightenings, which occurs between opposite polarity magnetic field patches (indicated with red and blue contours).
EUV brightenings are usually observed to occur between opposite magnetic field polarities\citep[e.g.][]{Zhukov2021, Kahil2022}. While it is beyond the scope of the current study to investigate the relationship between individual simulated brightenings and the magnetic field configuration, this will be considered in the future, when high resolution SO/PHI observations
are available to drive the simulation at the same time and with the same cadence as the EUI observations.

%

%
 \begin{figure} 
 \centerline{\includegraphics[width=1.0\textwidth,clip=]{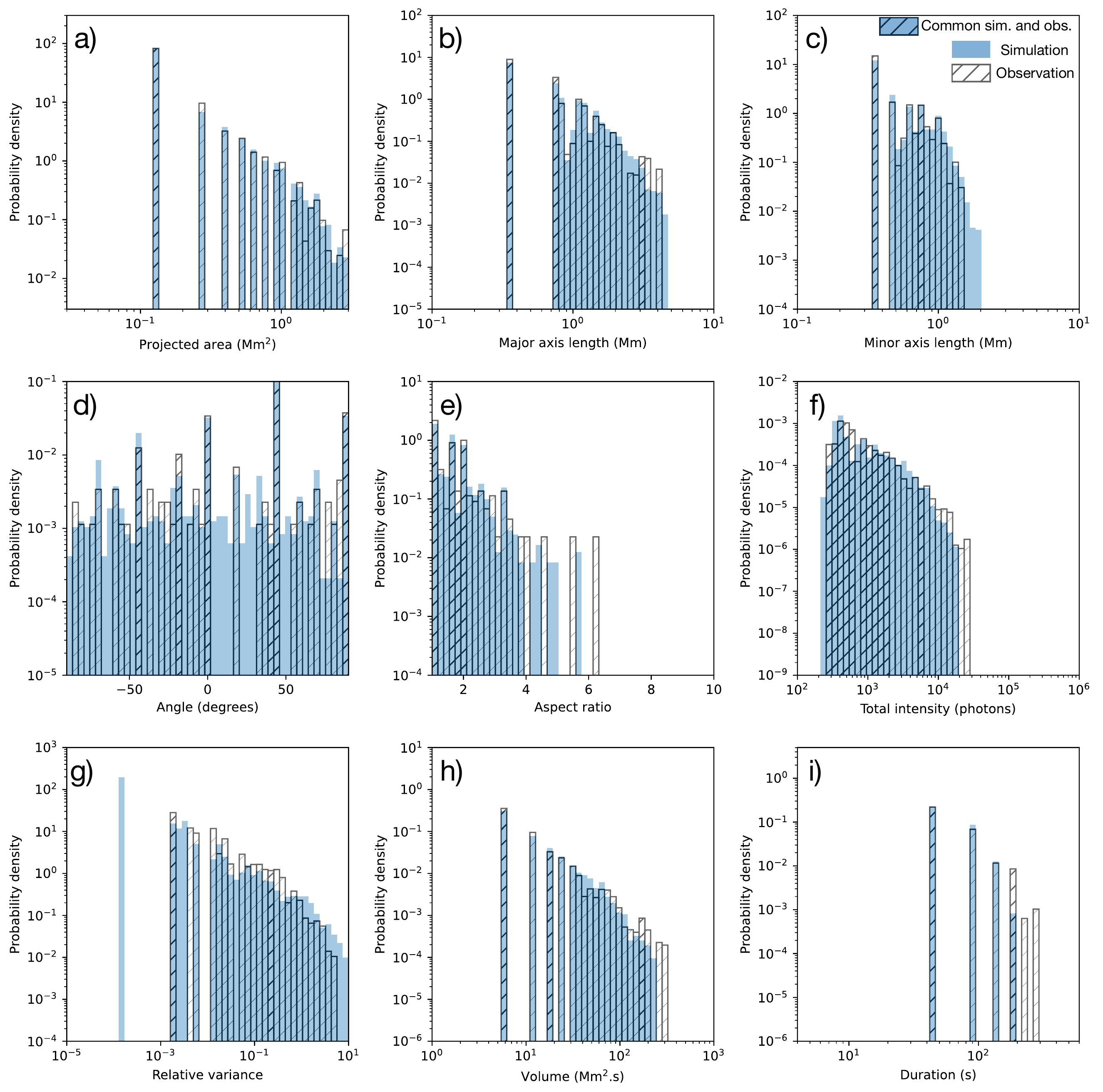}}
 \caption{The probability density of the physical and geometrical properties of EUV brightenings detected in EUI observation (hatch-filled) and simulated brightenings detected in the non-potential simulation (blue).
 The probability density distribution common for the simulation and the observation is hatch-filled with blue background.
 The histograms show distributions of the projected area, length (estimated as the size of the major axis of a fitted ellipse), width (estimated as the size of ellipse minor axis), orientation angle, length to the width aspect ratio, total intensity, relative variance of intensity, volume, and duration of the brightenings.
 The histograms are based on 240 EUV brightenings detected in EUI observations and 
 1310 simulated brightenings detected in the non-potential simulation. The probability density is scaled such that when integrated over all bin widths, the total event count is obtained.
 }\label{fig:sim_vs_obs}
 \end{figure}

\section{Comparison of Observed EUV Brightenings and Simulated Brightenings Properties}\label{s:cf_comp}

We compared the properties of the brightenings obtained from the observations and simulation. 
Figure~\ref{fig:sim_vs_obs} shows histograms of probability densities for various properties of the EUI observation (hatch-filled) and simulation (blue) brightenings.


The automatic detection method of \cite{Berghmans2021} produces an $(x,y,t)$ cube of events for both the observations and simulation. The projected area of a brightening is defined as the area of its projection along the temporal axis. The distributions of projected area (Figure~\ref{fig:sim_vs_obs}(a)) are visually very similar for both simulation and observation.

An ellipse is fitted to each brightening event, to give estimates of its length (ellipse major axis), width (ellipse minor axis) and orientation angle (of ellipse major axis). The distributions of ellipse major axis, minor axis, and angle for the observed and simulated brightenings can be seen in Figures~\ref{fig:sim_vs_obs}(b), (c), and (d), respectively. The observed and simulated distributions are visually very similar for both the major and minor axis length, particularly for smaller lengths. Only 0.38\% of the simulated brightenings have larger minor axis length than those detected in the observations.

The simulation and observation show an almost uniform distribution of the brightenings' orientation angle (Figure~\ref{fig:sim_vs_obs}(d)), with one exception. 
There appears to be a reduced number of EUV brightenings with a major axis angle between 0 and 30 degrees. This is likely due to the relatively small number of EUV brightenings considered (240), however, rather than a real phenomenon. 

The aspect ratio of the brightenings is calculated by dividing their length by their width (major axis length divided by the minor axis length of the fitted ellipse). The distributions of aspect ratios of observed and simulated brightenings are similar, for aspect ratios below 4. The difference in their distributions for brightenings with aspect ratios greater than 4 are related to only 1.7\% of the brightenings investigated.


The total intensity of a brightening is determined by integrating its intensity over its duration. The distributions of intensity for the observed and simulated brightenings are visually similar (Figure~\ref{fig:sim_vs_obs}(f)). Similar to before, only 1.25\% of the EUV brightenings were found to have intensities stronger than any detected in the simulation.

The relative variance of the intensity of a brightening is calculated as the variance of the mean intensity at each time, normalised to the mean intensity of the brightening throughout its duration. This was calculated only for brightenings that were observed in at least two images, corresponding to 68 (28\%) of the observed brightenings and 129 (10\%) of the simulated brightenings. The histogram in (Figure~\ref{fig:sim_vs_obs}(g)) shows that there exists simulated brightenings with both lower and higher relative variance in intensity than are detected in the observations. This corresponds to 1.5\% of simulated brightenings having lower relative intensity variance than any of the observed brightenings\footnote{1.5\% of simulated brightenings which exist in at least two frames, or 0.15\% of all simulated brightenings.}.
Similarly, 12\% of simulated brightenings having higher relative intensity variance than any of the observed brightenings\footnote{12\% of simulated brightenings which exist in at least two frames, or 1.2\% of all simulated brightenings.}.

The volume of a brightening is determined by the number of voxels associated with it in the $(x,y,t)$ cube of events, so has units of Mm$^{2}\cdot$s.
A visual inspection suggests almost the same distribution of the brightenings' volume (Figure~\ref{fig:sim_vs_obs}(h)) for both the simulation and observations.
The difference between the observation and simulation is related only to 1.25\% of the observed EUV brightenings,
which have a volume larger than 250 Mm$^{2}\cdot$s. Such brightenings are not detected in the simulated data.

Figure~\ref{fig:sim_vs_obs}(i) shows the distribution of brightening durations for the observations and simulation. The distributions are almost identical for brightenings of duration less than 150 s. Brightenings with a duration greater than 150~s are detected only in the observations and not in the simulation.

A statistical test was carried out to compare the distributions of brightening properties between the observations and simulation. Based on visual inspection and previous work \citep[e.g.][]{Alipour_2022}, we assumed that projected area, major axis, minor axis, aspect ratio, intensity, relative variance, volume, and duration of brightenings have a log-normal distribution. We assumed that the angle of the major axis has a uniform distribution. 

A statistical two-sample Kolmogorov-Smirnov (KS) test was used to compare the distributions of properties. This tests the assumption that each of the properties of the observed and simulated brightenings are drawn from the same distribution, i.e. the null hypothesis supposes no significant difference between the observed and simulated distributions of a brightening property. The alternate hypothesis supposes that there is a significant difference between the observed and simulation distributions for that property.

The p-value indicates whether the null hypothesis is plausible (p-value $>$ 0.05) or should be rejected (p-value $<$ 0.05). The p-values for the KS test applied to the observed and simulated distributions of each of the brightening properties are presented in Table~\ref{tbl:comp0}.

\begin{table}[ht!]
\begin{tabular}{@{}l|c|@{}}
Statistical parameter/EUI property   &  p-value \\ \midrule
a) Projected area   & 0.96   \\
b) Major axis       & 0.84  \\
c) Minor axis       & 0.99  \\
d) Angle (uniform dist.) & 0.07  \\
e) Aspect ratio     & 0.55   \\
f) Total Intensity  & 0.08  \\
g) Relative variance & 0.89  \\
h) Volume           & 0.27  \\
i) Duration         & \textless{}0.05  \\
 \bottomrule
\end{tabular}
\caption{The p-values resulting from the Kolmogorov-Smirnov test used to compare distributions of brightening properties obtained from observations and the simulation. A p-value $>$ 0.05 supposes no significant difference between the observational and simulated distributions.}\label{tbl:comp0}
\end{table}

The p-values for all properties except duration are greater than 0.05, indicating that we cannot reject the null hypothesis: namely that there is no significant difference between the observed and simulated distributions for those properties. The p-value for the duration of the brightenings is less than 0.05, however, suggesting that there is a significant difference between observed and simulated distributions for the duration.

The p-values for geometric properties tend to be high, e.g projected area, major axis and minor axis. The p-value for the distribution of angles (orientation) of the brightenings is low (0.07), although still $>$0.05. This may be due to the relatively small number of observed brightenings considered (240). As discussed above, there appear to be very few observed brightenings with major axis angle between 0 to 30 degrees (Figure~\ref{fig:sim_vs_obs}). In a future study, we will consider a larger area and longer duration of EUI observations, hence a much greater number of events will be sampled, to determine whether the distribution of angles is indeed uniform. 

The p-values for properties that depend on time are typically lower. This includes the duration, intensity and volume. In particular, the distributions of the duration of brightenings are based on only six frames; hence presenting discrete values. The KS test assumes that the two samples are drawn from the same \emph{continuous} distribution. Thus to improve the statistical comparison, the analysis of the duration of observed and simulated brightenings should be repeated for a simulation based on higher time cadence magnetic field observations (e.g. SO/PHI), to reduce the impact of the discrete nature of the data.

Four additional simulations were run to investigate the effect on the results of varying the simulation setup. The parameters varied were how long before the EUI observations the simulation was started; the inclusion of a diffusive term in the coronal magnetic field induction equation; and an open top boundary condition. The additional simulations are described in detail in Appendix~\ref{a:sim}, with detailed results presented in Appendix~\ref{a:res}. The total number of brightenings and the distribution of brightening properties are very similar between the original and additional simulations. The KS test was used to compare the distributions of brightening properties between each of the additional simulations and the observations. The p-values are presented in Table~\ref{tbl:comp1}. In the additional simulations we also find that we  cannot reject the null hypothesis for the majority of the properties: that the observed and simulated brightening properties are drawn from the same distribution. The only exceptions are for the brightening duration, and for some simulations, the brightening total intensity (which is calculated by integrating over the duration). This indicates that to investigate it further the analysis should be repeated with higher temporal resolution data.

The brightening number density is 26.8 times larger in the non-potential simulation compared to the observations.
The simulated images were produced by integrating $Q$ (see Section~\ref{s:cmodel}) from z$_{\rm min}$ to z$_{\rm max}$, i.e. from the photosphere to the top boundary of the simulation.
One can consider that occurrences of energy release closer to the photosphere may not be observed due to absorption by the denser plasma.
$Q$ decreases quite rapidly with height in the simulation.
Most EUV brightenings are observed between 1000 – 5000 km \citep{Zhukov2021}. 
Due to this, we tested the energy dissipation based upon an increased height of the lower bound (z$_{\rm min}$) for the line-of-sight integration. When we set z$_{\rm min}$ at 364 km (1 grid point),
728 km (2 grid points) or 1092 km (3 grid points) we obtain 1431, 1248, and 955 simulated brightenings respectively, which corresponds to density $4.14\times 10^{-2}$ brightenings/Mm$^{2}$, $3.61\times 10^{-2}$ brightenings/Mm$^{2}$ and $2.76\times 10^{-2}$ brightenings/Mm$^{2}$ respectively.

%
%
%
%
Thus, the brightenings density detected in the simulation decreases with increasing z$_{\rm min}$ above z$_{\rm min}=364$ km.
However, even with this decreased number the simulated brightenings, the density is still 19.57 times larger in the non-potential simulations than in the observations, for the simulation integrating $Q$ from z$_{\rm min}=1092$ km upwards.

The simulated brightening density may be larger than the observed one because the simulation does not reproduce the detailed thermal properties of the atmosphere.
The cooling, heating, and energy deposition can significantly influence the properties and lifetime of the small-scale features.

In a future study, we will consider properties of simulated brightenings at specific locations. This will include specific structures in the magnetic configuration and 
the twist, $\alpha$, of the non-potential field. This will allow us to investigate whether there is a relationship between such properties and the simulated brightening being colocated with an observed brightening.

\section{Conclusion}\label{s:conclusion}

We compared the properties of the EUV brightenings observed in EUI and simulated brightenings obtained from a non-potential coronal magnetic field simulation.
The automatic EUV brightenings detection method of \cite{Berghmans2021} used originally 
with EUI data can also be applied to simulated data.
We detected 240 EUV brightenings in EUI images and between 955-1431 simulated brightenings in simulated images. The number obtained from the simulations depends on the vertical extent over which the energy distribution is computed in the non-potential simulations.
The brightenings detected in the EUI images and simulated images show very similar distributions of the basic geometrical properties (projected area, volume, length, width, length to width aspect ratio, angle) and to a lesser extent, physical properties (total intensity, intensity variation).
Moreover, the lower and the upper limits of the geometrical and physical properties are very similar for the brightenings detected in EUI observations and the simulation.
The lower limit of each distribution is a result of using the same spatial and temporal resolution to compare the simulation and observations (having re-sampled the HRI$_{\rm EUV}$ observations to match the HMI resolutions).
The upper limit is determined by the wavelet scales in the detection algorithm.
The visual similarities in distributions are backed up by performing a statistical two-sample Kolmogorov-Smirnov test to compare the observed and simulated distributions of each of the brightening properties. It was found that we cannot reject the null hypothesis, that there is no significant difference between the observed and simulated distributions, for all brightening properties except the duration. Other properties that depend on time also had lower p-values, although not low enough to indicate a significant difference between the distributions.

The analysis here was carried out on only 6 frames of observed and simulated data, resulting in few discrete values for duration. Moreover, the total number of observed brightenings (240) in the study is relatively low. The analysis should be repeated with simulations driven by higher cadence magnetograms (e.g. SO/PHI), so that a larger number of frames and larger number of observed brightenings can be compared.

The similarities in distributions of the basic geometrical and physical properties suggests that the simulated energy dissipation with the method presented by \citet{Meyer2012} and
\citet{Meyer2013} reproduces the EUV brightenings with their basic geometrical and physical properties.

While the observations and simulations show a high level of agreement in terms of several statistical properties (geometrical and physical) there is however one important difference.
The number density of brightenings detected in the simulation is between 19-29 times larger than the number density obtained in the observations.
It is important to consider why the non-potential simulations produce a significantly higher number density of the brightenings compared to the observations. While there may be a number of reasons, one of the most important will be the simplicity of the thermodynamics in the simulation, which produces the energy dissipation in the solar atmosphere based on the magnetofrictional method.
With this method the simulated images are constructed through simply integrating magnetic energy dissipation along the LOS. This is an over-simplification, but to improve on this, radiative forward modelling would need to be carried out.
In addition, the simulations do not take into account the thermal properties of the solar atmosphere, such as radiative losses and thermal conduction.
Such thermodynamic processes in the solar atmosphere reduces the EUV brightenings evolution dynamics by removing and redistributing energy, which would shorten the brightenings' lifetimes.
Thus, considering the impact of the thermodynamic processes, if included we would expect a smaller number of brightenings to be detected in the simulations. Developing such a model based on driving through observed magnetograms is beyond the scope of the present study, but may be considered in future studies.

The simulation gives a unique opportunity to study the dynamics, geometrical and physical properties of brightenings. We have shown that basic geometrical and physical properties of EUV brightenings are consistent with the excess energy that is released as the corona evolves through a series of non-linear force-free states.
In a further study, the magnetofrictional simulation method will be used with high-resolution magnetic field data obtained with Polarimetric and Helioseismic Imager (PHI), especially with data obtained during Solar Orbiter perihelia, as this will allow the data driven model to simulate the same length scales and timescales as seen in the observations. 
During the last Solar Orbiter perihelion (26 March 2022) the pixel size corresponds to 116 km.
Moreover, the Daniel K. Inouye Solar Telescope (DKIST) should also be used to observe the EUV brightenings.

%

%
\appendix 
\section{Additional Simulations}
\subsection{Simulation Setup}\label{a:sim}
Four additional simulations were run to test the effect of varying the simulation setup on the results. The simulation described in Section~\ref{s:cmodel} was initiated 6 hours before the EUI observations, to allow sufficient time for the coronal field to evolve away from its potential field initial condition. The timescale for the simulated coronal magnetic field to evolve to a self-consistent state is based on the applied surface motions from the observed magnetograms, which include surface motions due to granulation and supergranulation, as well as the emergence, cancellation, coalescence and fragmentation of magnetic flux. The timescale for the coronal field to reach a self-consistent state is therefore determined by the photospheric flux replacement time, which is of order $1-2$ hours \citep{Hagenaar2008}. Hence, initiating the simulation from HMI observations 6 hours before the EUI observations was determined to be more than sufficient.

To investigate the effect of varying the simulation start time, two additional simulations were run, starting 4 hours and 2 hours before the EUI observations, at 10:57:14 UTC and 12:57:14 UTC, respectively. In the results section below, we will refer to the simulations starting 2, 4 and 6 hours before the EUI observations as `2h', `4h' and `6h'.

The third additional simulation to be run included a diffusive term in the coronal magnetic field induction equation, so that Equation~(\ref{eqn:ind}) becomes:
$$\frac{\partial\mathbf{A}}{\partial t}=\mathbf{v}\times\mathbf{B}+\textrm{\boldmath$\epsilon$}.$$
The term ${\boldmath\epsilon}$ represents \emph{hyperdiffusion}:
$$\textrm{\boldmath$\epsilon$}=\frac{\mathbf{B}}{B^2}\nabla\cdot(\eta_4B^2\nabla\alpha),$$
where
$$\alpha=\frac{\mathbf{j}\cdot\mathbf{B}}{B^2}$$
describes the twist of the magnetic field with respect to the corresponding potential field. Hyperdiffusion acts to smooth gradients in $\alpha$ while conserving magnetic helicity. For this simulation, we take $\eta_4=1.9\times10^5$ km$^4$ s$^{-1}$. All other aspects of the simulation setup are the same as in the 2 hour simulation, `2h', described above.

Due to the additional term in the coronal field induction equation, Equation~(\ref{eqn:q}) becomes:
\begin{equation}\label{eqn:qhd}
    Q=\frac{B^2}{4\pi}(\nu|\mathbf{v}|^2+\eta_4|\nabla\alpha|^2).
\end{equation}
See e.g. \cite{Meyer2012,Meyer2013} for further details. $Q$ is integrated along the line of sight using Equation~(\ref{eqn:qxy}) to determine $E_q(x,y)$, as in the other simulations. This simulation will be referred to as `2h\_hd' in the results section below.

The fourth additional simulation to be run has the same setup as the 2 hour simulation, `2h', described above, but with an open top boundary and without applying the correction for flux imbalance described in Section~\ref{s:hmi}. The region is relatively close to flux balance in any case, with an average and maximum imbalance per magnetogram of 2.3 \% and 4 \%, respectively. This simulation will be referred to as `2h\_open' in the results section below.

\subsection{Results of Additional Simulations}\label{a:res}
\begin{figure}[ht!]
 \centerline{\includegraphics[width=1.0\textwidth,clip=]{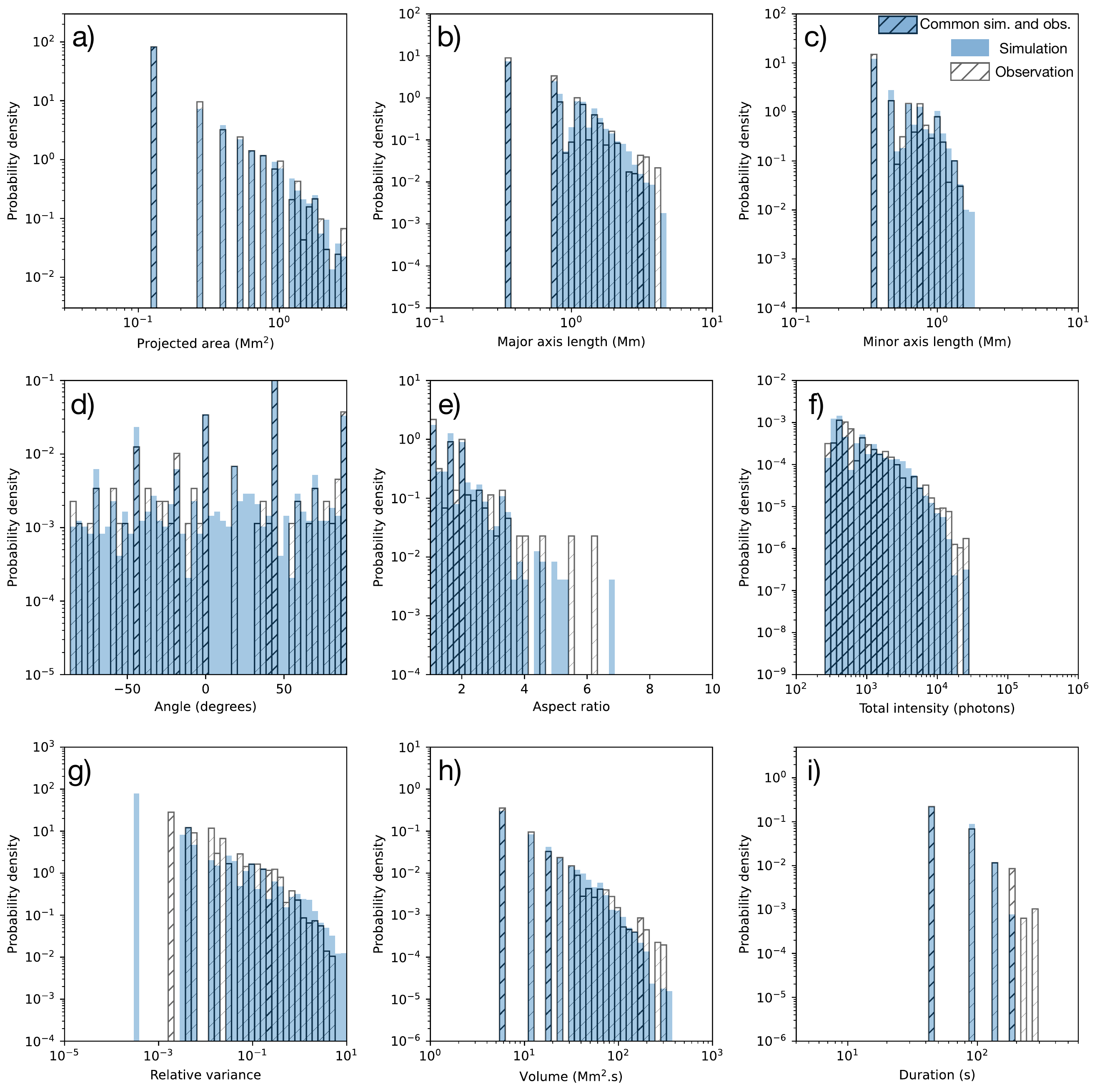}}
 \caption{The probability density of the physical and geometrical properties of EUV brightenings detected in EUI observations (hatch-filled) and simulated brightenings detected in the non-potential simulation, `4h' (blue). The simulation begins four hours before the observations. The probability density distribution common for the simulation and the observation is hatch-filled with blue background. The histograms show distributions of the projected area, length (estimated as the size of the major axis of a fitted ellipse), width (estimated as the size of ellipse minor axis), orientation angle, length to the width aspect ratio, total intensity, relative variance of intensity, volume, and duration of the brightenings.
 The histograms are based on 240 EUV brightenings detected in EUI observations and 1313 simulated brightenings detected in the non-potential simulation. The probability density is scaled such that when integrated over all bin widths, the total event count is obtained.
 }\label{fig:sim_vs_obs_4h}
 \end{figure}

\begin{figure}[ht!]
 \centerline{\includegraphics[width=1.0\textwidth,clip=]{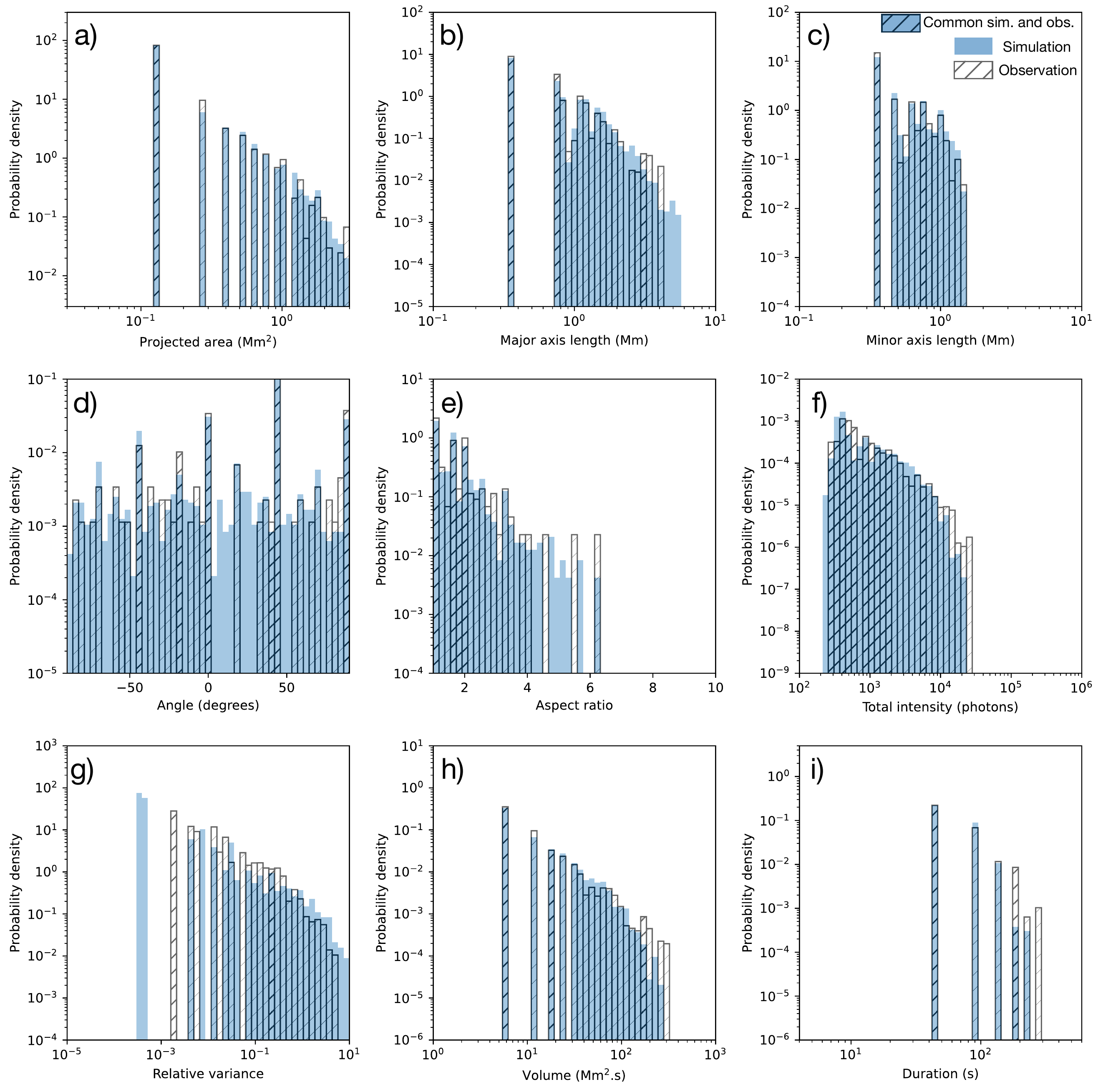}}
 \caption{The probability density of the physical and geometrical properties of EUV brightenings detected in EUI observations (hatch-filled) and simulated brightenings detected in the non-potential simulation, `2h' (blue). The simulation begins two hours before the observations. The probability density distribution common for the simulation and the observation is hatch-filled with blue background. The histograms show distributions of the projected area, length (estimated as the size of the major axis of a fitted ellipse), width (estimated as the size of ellipse minor axis), orientation angle, length to the width aspect ratio, total intensity, relative variance of intensity, volume, and duration of the brightenings.
 The histograms are based on 240 EUV brightenings detected in EUI observations and 1301 simulated brightenings detected in the non-potential simulation. The probability density is scaled such that when integrated over all bin widths, the total event count is obtained.
 }\label{fig:sim_vs_obs_2h}
 \end{figure}

\begin{figure}[ht!]
 \centerline{\includegraphics[width=1.0\textwidth,clip=]{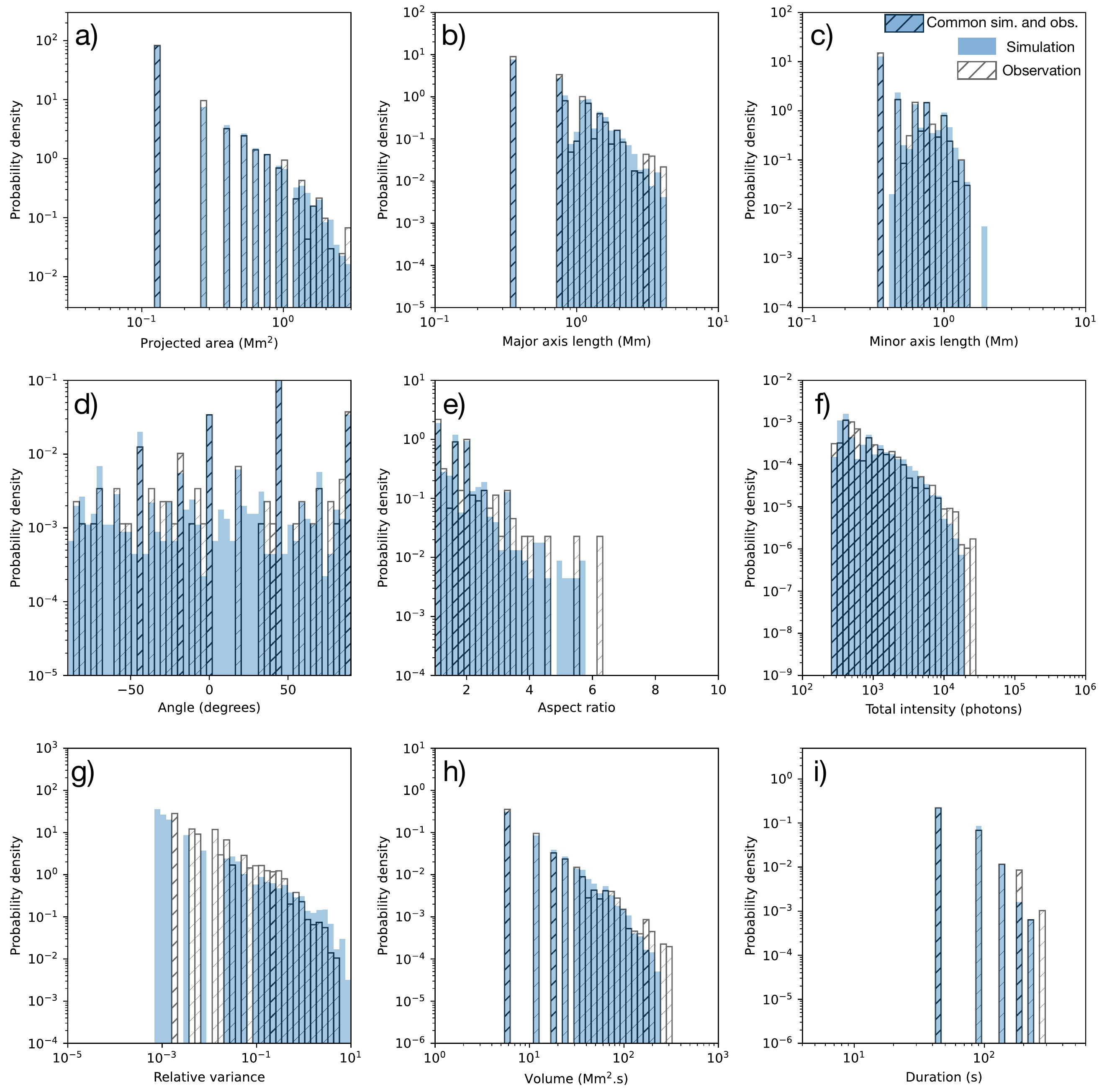}}
 \caption{The probability density of the physical and geometrical properties of EUV brightenings detected in EUI observations (hatch-filled) and simulated brightenings detected in the non-potential simulation with hyperdiffusion, `2h\_hd' (blue). The simulation begins two hours before the observations. The probability density distribution common for the simulation and the observation is hatch-filled with blue background.  The histograms show distributions of the projected area, length (estimated as the size of the major axis of a fitted ellipse), width (estimated as the size of ellipse minor axis), orientation angle, length to the width aspect ratio, total intensity, relative variance of intensity, volume, and duration of the brightenings.
 The histograms are based on 240 EUV brightenings detected in EUI observations and 1233 simulated brightenings detected in the non-potential simulation. The probability density is scaled such that when integrated over all bin widths, the total event count is obtained.}
 \label{fig:sim_vs_obs_2h_hiperdiff}
 \end{figure}

\begin{figure}[ht!]
 \centerline{\includegraphics[width=1.0\textwidth,clip=]{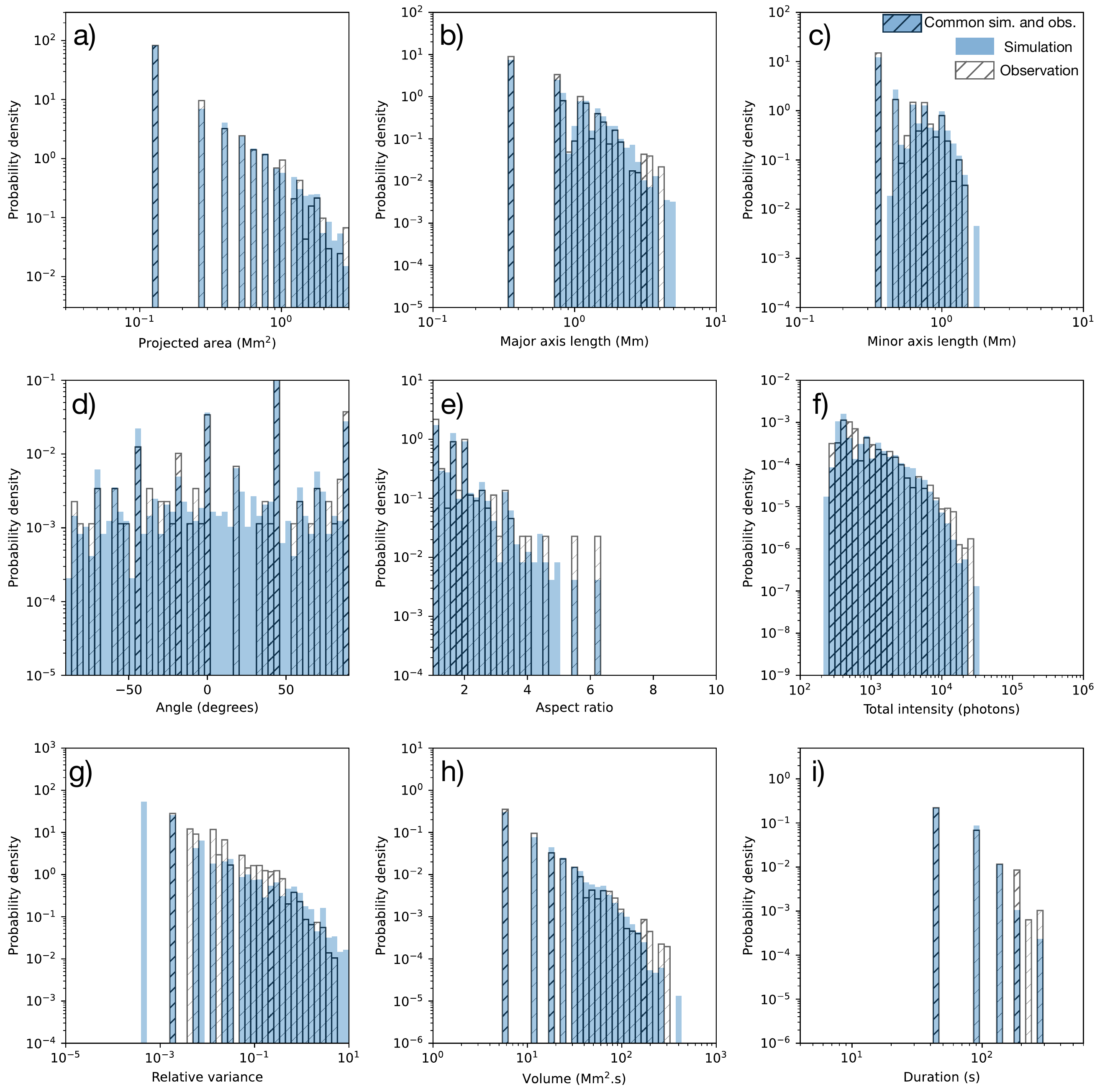}}
 \caption{The probability density of the physical and geometrical properties of EUV brightenings detected in EUI observation (hatch-filled) and simulated brightenings detected in the non-potential simulation with open top boundary, `2h\_open' (blue). The simulation begins two hours before the observations. The probability density distribution common for the simulation and the observation is hatch-filled with blue background. The histograms show distributions of the projected area, length (estimated as the size of the major axis of a fitted ellipse), width (estimated as the size of ellipse minor axis), orientation angle, length to the width aspect ratio, total intensity, relative variance of intensity, volume, and duration of the brightenings.
 The histograms are based on 240 EUV brightenings detected in EUI observations and 1318 simulated brightenings detected in the non-potential simulation. The probability density is scaled such that when integrated over all bin widths, the total event count is obtained.
 }\label{fig:sim_vs_obs_2h_open}
 \end{figure}

We computed the distribution of the properties of simulated brightenings for the four additional simulations. Probability density plots are presented for the same nine properties considered in Section~\ref{s:cf_comp} (`6h' simulation), for the simulations `4h' (Figure~\ref{fig:sim_vs_obs_4h}), `2h' (Figure~\ref{fig:sim_vs_obs_2h}), `2h\_hd' (Figure~\ref{fig:sim_vs_obs_2h_hiperdiff}) and `2h\_open' (Figure~\ref{fig:sim_vs_obs_2h_open}). In each case, the probability density of the simulated brightenings is plotted in blue, with the probability density of the observed EUV brightenings plotted as hatch-filled bars. Comparing Figures~\ref{fig:sim_vs_obs_4h}--\ref{fig:sim_vs_obs_2h_open} and the `6h' case (Figure~\ref{fig:sim_vs_obs}), the probability densities for each property appear to be very similar for all simulations.
To quantify this, we compared distribution of brightening properties obtained from the observation and simulations using a Kolmogorov-Smirnov test, in the same manner as in Section~\ref{s:cf_comp}.
The p-values obtained for all simulations tested against the observations are summarised in Table~\ref{tbl:comp1}. The results of the four additional simulations are very similar to those found for the `6h' simulation discussed in Section~\ref{s:cf_comp}. High p-values are typically found for geometrical properties such as projected area, major axis, and minor axis, indicating that we cannot reject the null hypothesis: that the observed and simulated brightening properties have the same distribution. Properties that depend on time, such as duration, volume, and intensity typically have lower p-values. The p-values are less than 0.05 for the duration of brightenings for all simulations, and for the intensity in two cases (`2h' and `2h\_open'), indicating that there is a significant different between the distributions of simulated and observed brightenings for these properties. It should be noted that these results are based on only 6 frames of data, however. These properties should be investigated in the future using a longer dataset and high cadence SO/PHI magnetogram observations to drive the simulations, so that a much larger number of frames can be considered.

The total number of brightenings detected in each simulation are similar, with slightly fewer in the simulation with hyperdiffusion (5.9\%). The total number of brightenings for each simulation and the observations are presented in Table~\ref{tbl:num}, along with the number density of brightenings. Note that the area considered by the simulations is smaller than the area considered in the observations, so the density of brightenings should be compared between them rather than the total number.

\begin{table}[ht!]
\begin{tabular}{@{}l|r|r|r|r|r|@{}}
 & \multicolumn{5}{c|}{Simulation name} \\ \midrule
\begin{tabular}[c]{@{}l@{}}Statistical parameter\\ /EUI property\end{tabular}   &  6h & 4h & 2h & 2h\_hd & 2h\_open \\ \midrule
a) Projected area   & 0.96 & 0.95 & 0.95 & 0.94 & 0.94  \\
b) Major axis       & 0.84 & 0.80 & 0.85 & 0.84 & 0.82 \\
c) Minor axis       & 0.99 & 0.99 & 0.99 & 0.99 & 0.99 \\
d) Angle (uniform dist.) & 0.07 & 0.05 & 0.08 & 0.14 & 0.02\\
e) Aspect ratio     & 0.55 & 0.61 & 0.33 & 0.58 & 0.77 \\
f) Intensity        & 0.08 & 0.05 & \textless{}0.05 & \textless{}0.05 & 0.07 \\
g) Relative variance & 0.89 & 0.94 & 0.97 & 0.94 & 0.99 \\
h) Volume           & 0.27 & 0.22 & 0.40 & 0.34 & 0.48 \\
i) Duration         & \textless{}0.05 & \textless{}0.05 & \textless{}0.05 & \textless{}0.05 & \textless{}0.05  \\\bottomrule
\end{tabular}
\caption{The p-values resulting from the Kolmogorov-Smirnov test used to compare distributions of brightening properties obtained from observations and the five simulation cases. The simulations started 6 (`6h'), 4 (`4h') and 2 (`2h') hours before the EUV brightenings observation. Two additional 2 hour simulations include one with hyperdiffusion (`2h\_hd') and one with an open top boundary (`2h\_open'). A p-value $>$ 0.05 supposes no significant difference between the observational and simulated distributions.}\label{tbl:comp1}
\end{table}

\begin{table}[ht!]
\begin{tabular}{@{}l|r|r|r|r|r||r|@{}}
 Simulation name  &  6h & 4h & 2h & 2h\_hd & 2h\_open & observed \\ \midrule
Total number of brightenings   & 1310 & 1313 & 1301 & 1233 & 1318 & 240  \\\midrule
Density of brightenings      & 3.79 & 3.80 & 3.76 & 3.56 & 3.81 & 0.14 \\
($\times10^{-2}$ brightenings/Mm$^2$) & & & & & & \\\bottomrule
\end{tabular}
\caption{Total number of brightenings detected and number density of brightenings for each simulation and for the EUI observations. Note that the observations and simulations consider different areas, so the number density can be used to compare directly between them.}\label{tbl:num}
\end{table}

%
 \begin{acks}
 Solar Orbiter is a space mission of international collaboration between ESA and NASA, operated by ESA. The EUI instrument was built by CSL, IAS, MPS, MSSL/UCL, PMOD/WRC, ROB, LCF/IO with funding from the Belgian Federal Science Policy Office (BELPSO); the Centre National d’Etudes Spatiales (CNES); the UK Space Agency (UKSA); the Bundesministerium für Wirtschaft und Energie (BMWi) through the Deutsches Zentrum für Luft- und Raumfahrt (DLR); and the Swiss Space Office (SSO). We thank the anonymous reviewer for careful reading of our manuscript and many insightful comments and suggestions.
 
 LH and KB are grateful to the SNF for the funding of the project number 200021\_188390.
DHM would like to thank the STFC for support via consolidated grant
ST/W001195/1. KAM would like to thank the STFC for support via consortium grant
ST/W001098/1.
 \end{acks}


%
%
 \bibliographystyle{spr-mp-sola}
 \bibliography{campfires_sym}
%
%
%
%

\end{article} 
\end{document}